\documentclass[12pt]{article}
\usepackage{amsfonts}
\usepackage{mathtools}
\usepackage{slashed}

\def\beq{\begin{equation}}
\def\eeq{\end{equation}}
\def\bea{\begin{eqnarray}}
\def\eea{\end{eqnarray}}
\def\ba{\begin{array}}
\def\ea{\end{array}}
\def\pa{\partial}

\def\nn{\nonumber}

\def\Ga{\Gamma}
\def\tGa{\widetilde\Gamma}

\def\de{\delta}

\def\la{\lambda}

\def\vp{\varphi}
\def\cD{{\cal{D}}}
\def\cO{{\cal O}}
\def\cL{{\cal L}}

\def\rd{{\rm d}}

\def\rd{{\rm d}}

\begin{document}
\hfill AEI-2013-159\\

\centerline{\bf\Large The DeWitt Equation in Quantum Field Theory}

\vspace{5mm}
\centerline{\bf Parikshit Dutta${}^1$, Krzysztof A. Meissner${}^2$ and Hermann Nicolai${}^1$}

\vspace{5mm}
\begin{center}
{\it ${}^1$ Max-Planck-Institut
f\"ur Gravitationsphysik (Albert-Einstein-Institut),\\
M\"uhlenberg 1, D-14476 Potsdam, Germany\\ ${}^2$ Institute of Theoretical Physics, Faculty of Physics,\\
University of Warsaw, Ho\.za 69, 00-681 Warsaw, Poland}
\end{center}

\begin{abstract}
\footnotesize{We take a new look at the DeWitt  equation, a defining equation for the
effective action functional in quantum field theory. We present a formal solution to this
equation, and discuss the equation in various contexts, and in particular for models
where it can be made completely well defined, such as the Wess-Zumino model in two dimensions.}
\end{abstract}

\section{Introduction}

In 1965, B. DeWitt wrote down a functional differential equation for the full effective
action in quantum field theory \cite{deWitt}. To the best of our knowledge,
this result has not received much attention in the existing literature (but see 
\cite{FM} for a recent exception), nor in quantum field theory textbooks.
The equation in question,
which in the remainder we will refer to as the `DeWitt equation'~\footnote{Not
 to be confused with the more famous Wheeler-DeWitt equation!}, relates the functional
derivative of the full quantum effective action $\Gamma[\vp]$ to the functional derivative 
of the classical action, and has several remarkable features. First of all, while the usual
approach to quantum field theory is based on path integrals and perturbation theory, and thus involves (functional) {\em integration} (see e.g. \cite{GJ}), the essential information about 
the quantum field theory is here encoded into a (functional) {\em differential equation}.
If the classical action is polynomial, this equation has only very few terms
and therefore assumes a relatively simple form. Second, this
equation can serve as the generating equation for an infinite hierarchy of Schwinger-Dyson equations for the theory in question. 

The main difficulty, and possibly the reason why this equation has not 
been much exploited in the past, is that it is even hard to {\em define} properly.
Of course, this is also true of the path integral, but there one has a number of 
established approximation methods at one's disposal (such as renormalized 
perturbation theory), whereas apparently no techniques exist as yet for
dealing with a functional differential equation that should contain the complete
information about the full renormalized action functional. Amongst other difficulties,
one has to deal with short distance singularities in the equation
related to the occurrence of functional derivatives at coincident points
that would have to be resolved `in one stroke', rather than by perturbative
methods of conventional type.
Consequently the proper definition of the equation already requires some 
knowledge of the properties of the solution. One possible approach here would be to
to look for formal solutions in a perturbative expansion of the unrenormalized equation, and then renormalize the resulting expression in a second step \cite{FM}. A second difficulty is 
that the equation is not of any known type, even in a discrete approximation 
with only finitely many degrees of freedom (which we consider in section~3).

In this paper we take a new look at the DeWitt equation, and will argue that, 
in spite of the difficulties mentioned above, the equation may provide valuable new insights 
into quantum field theory, beyond the established results and techniques used so far. 
Our main motivation here is to be able eventually to
develop new methods for future applications, in order
to deal with the effective (Coleman-Weinberg) potential \cite{CW} in 
classically conformal versions of the Standard Model 
of the type considered in \cite{MN}, possessing more than one scalar degree of freedom.
As argued there (see also \cite{Bardeen,MN1}), classically unbroken conformal 
symmetry may offer an  attractive alternative to low energy supersymmetry 
in explaining the stability of the electroweak scale. 
The main technical problem with this proposal is that, so far, there 
appear to be no efficient methods to compute the effective potential with more than one
physical scalar field beyond one loop. However, such methods are absolutely required 
in order to reliably assess the existence and stability of non-trivial stationary points of
the effective potential because the extremal structure of the potential may 
delicately depend on higher order corrections.

In fact, as we will show, there exists a formal solution to the DeWitt equation, which
represents the effective action functional $\Ga[\vp]$ as an asymptotic series expansion 
over vacuum diagrams with field-dependent Green's functions; this result follows
from much older results on the effective potential obtained by R.~Jackiw \cite{Jackiw}
(see also \cite{Jackiw2}). One interesting new aspect here is that this analysis leads
us to consider the question of convergence of such expansions not only in terms
of the coupling constants (or running coupling constants), but rather as a question
of convergence {\em in field space}\,: the value of the classical field $\vp$ effectively 
replaces the renormalization scale of the usual perturbation expansion. In this case, 
Landau poles and other singularities would manifest themselves as singularities
of the effective action in field space, while the couplings are kept fixed
and do not run.~\footnote{A standard example for this phenomenon 
is the RG improved  effective potential for $\phi^4$ theory, which takes the form 
$$
V_{\rm eff} (\vp) = \frac{\la\vp^4}{1 - a\la \log(\vp^2/v^2)}
$$
where $a$ is a positive constant and $v$ a fixed scale \cite{CW}. This expression is
thus valid only in a compact region in $\vp$-space. See also \cite{MN2} for 
further examples and a discussion of this issue in the case with one
scalar field, but any number of non-scalar fields.}
As we will show,  also in terms of explicit numerical examples
(see appendix), the new expansion may have much better convergence properties
even for large coupling constants $\la$, if the value of the classical field $\vp$
is different from zero.

Another new direction opened by this work concerns the 
formulation of the DeWitt equation in contexts where it can be made completely 
well defined. Our prime example here is the Wess-Zumino model
in two space-time dimensions, where we can exploit the cancellation of UV singularities
in a supersymmetric model. As a further application, we derive the DeWitt equation
for Liouville theory in two dimensions, as an example of a theory with non-polynomial
action. In this way we are led to a novel relation between $n$-point correlators and
 $(n+1)$-point correlators of exponential Liouville operators, a (formal) result that
 that does not rely on conformal symmetry, and remains to be exploited in future work.
A most interesting future application of the present work would be the
formulation and analysis of the DeWitt equation for $N=4$ super-Yang-Mills theory,
the main example of a UV finite interacting quantum field theory in four space-time
dimensions \cite{olive,Brink}.

\section{Derivation of DeWitt Equation}

For the reader's convenience we here reproduce the formal derivation of
the equation found by B.~DeWitt, following \cite{deWitt} (see also \cite{FM}), restricting
attention to scalar field theory for simplicity, as the extension to more general theories 
(with fermions and gauge fields) is straightforward, at least  in principle.  
Working with a Euclidean metric for simplicity,  we define
the generating functional of connected Green's functions $W[J]$
in the standard way via (see e.g. \cite{IZ,Weinberg})
\beq\label{W}
Z[J] \equiv \exp \left[-\frac{1}{\hbar}W[J]\right] =
\int \cD \phi \exp\left[ -\frac{1}{\hbar}\Big( S[\phi] + J\cdot\phi\Big)
\right]
\eeq
where $J\cdot\phi\equiv \int dx J(x)\phi(x)$ and the measure $\cD\phi$
is formally normalized to unity, that is $W[0]=0$. The connected Green's
functions in the presence of a source $J$ are then given by
\beq\label{Wn0}
W_n (x_1,...,x_n;J) \equiv (-\hbar)^{n-1}
\frac{\de^n W[J]}{\de J(x_1) \cdots \de J(x_n)}\,
\eeq
with the full connected $n$-point functions
\beq\label{Wn}
W_n(x_1,...,x_n)\equiv W_n(x_1,...,x_n;J) \Big|_{J=0}
\eeq
Defining the classical field $\vp (x)$ by
\beq\label{vp}
\vp (x) \equiv \vp(x;J) = \frac{\de W[J]}{\de J(x)}
\eeq
the effective action is the Legendre transform
\beq\label{Gamma}
\Ga[\vp] = W[J] - \int d^4x \, J(x) \vp (x)
\eeq
such that
\beq
\frac{\de \Ga[\vp]}{\de \vp(x)}=-J(x;\vp)
\label{dgj}
\eeq
We will assume in the following that the relation between
$J=J(x;\vp)$ and $\vp(x;J)$ can be freely inverted (although we are
aware that this may not be true in many cases of physical interest!).
As is well known, $\Ga[\vp]$ is the generating functional for the one-particle
irreducible ($\equiv$ 1PI) Green's functions, with
\beq
\Ga_n(x_1,...,x_n;\vp) \equiv \frac{(-1)^{n}}{\hbar}\,
\frac{\de^n \Ga[\vp]}{\de \vp(x_1) \cdots \de \vp(x_n)}\,
\eeq
where the normalization is chosen such that we
have the standard relations
\beq
\int \rd^4y \, W_2(x,y;\vp) \Ga_2(y,z;\vp)  = \de^{(4)}(x-z)
\label{gga}
\eeq
and
\bea\label{W3}
W_3(x,y,z;\vp)   &=&  \\[1mm]
&& \!\!\!\!\!\! \!\!\!\!\!\! \!\!\!\!\!\!\!\!\!\!\!\! \!\!\!\!\!\!\!\!\!\!\!\!
 = \,  \int \rd^4u\,\rd^4v\, \rd^4w
\,W_2(x,u;\vp) W_2(y,v;\vp)W_2(z,w;\vp) \Ga_3(u,v,w;\vp) \nn
\eea
and so on. Note that here all the Green's functions depend on the
classical field $\vp(x)$. We also recall the expansion of the
effective action in powers of $\hbar$ (`loop expansion')
\beq\label{GaLoop}
\Ga[\vp]=\Ga^{(0)}[\vp]+\hbar\Ga^{(1)}[\vp]  +  \ldots
\eeq
where $\Ga^{(0)}[\vp]=S[\vp]$ is the classical action $S$.

For any functional $Q[\phi]$ we define the expectation value with given
source $J(x)$ as
\beq
\left\langle Q[\phi]\right\rangle_J := \exp\left(  \frac{1}{\hbar} W[J]\right)
\int \cD \phi Q[\phi]  \exp\left[ -\frac{1}{\hbar}\Big( S[\phi] + J\cdot\phi\Big) \right]
\eeq
This can be rewritten as~\footnote{By use of the elementary identity
$f(x) = \exp (x \pa/\pa y) f(y) |_{y=0}$.}
\beq\label{<QPhi>}
\left\langle Q[\phi]\right\rangle_J =  \exp\left( \frac{1}{\hbar} W[J]\right)
\exp\left( -\frac{1}{\hbar}W\left[ J - \hbar  \frac{\de}{\de\phi} \right]
\right)   Q[\phi] \Big|_{\phi =0}
\eeq
Next we expand
\bea
W\left[J -  \hbar \frac{\de}{\de\phi} \right] &=& W[J]
- \hbar \int \rd^4 x \, \frac{\de W[J]}{\de J(x)} \, \frac{\de}{\de\phi(x)} +
\\
&& \!\!\!\!\!\!\!\!\!\!\!\!\!\!\!\!\!\!\!\!\!\!
- \, \hbar \sum_{n=2}^\infty \frac1{n!} \int \rd^4x_1\cdots \rd^4 x_n
\,
W_n (x_1,...,x_n;J)\,
\frac{\de}{\de \phi(x_1)} \cdots \frac{\de}{\de \phi(x_n)} \nn
\eea
Expressing $J$ as a functional of $\vp$, using (\ref{vp}) and
once again the elementary identity from footnote~2 to replace
$\phi$ by $\vp$ in (\ref{<QPhi>}) we arrive at
\bea\label{Qvp}
\left\langle Q[\phi]\right\rangle_{J[\vp]} &=& \\[2mm]
&&  \!\!\!\!\!\!\!\!\!\!\!\!\!\!\!\!\!\!\!\!\!\!\!\!\!\!\!\!\!\!\!\!\!\!\!
{}_{*}^{*}\exp \left[ \sum_{n=2}^\infty \frac1{n!} \int \rd^4x_1\cdots
\rd^4 x_n \, W_n \Big(x_1,...,x_n;J[\vp]\Big)\,
\frac{\de}{\de \vp(x_1)} \cdots \frac{\de}{\de \vp(x_n)}\right]{}_{*}^{*} \; Q[\vp] \nn
\eea
where the symbol ${}_{*}^{*}$ indicates that the functional differential operators
act only on the external factor $Q[\vp]$, but not on $J[\vp]$ in $G_n$.
It is important here that the sum in the exponent starts only at $n=2$.
Next recall DeWitt's identity
\beq\label{GS}
\frac{\de\Ga[\vp]}{\de\vp(x)} =
\left\langle \frac{\de S[\phi]}{\de\phi(x)} \right\rangle_{J=J[\vp]}
\eeq
which holds since both sides are equal to $-J(x)$ (a consequence of
the formal identity $\int \cD \phi \, \delta/\delta\phi(x) (\cdots) = 0$).
DeWitt's equation is now obtained by applying (\ref{Qvp}) with
$Q[\phi]= \de S/\de \phi$.  This gives
\bea\label{Gvp}
\frac{\de\Ga[\vp]}{\de\vp(x)} &=&  \\[2mm]
&&  \!\!\!\!\!\!\!\!\!\!\!\!\!\!\!\!\!\!\!\!\!\!\!\!\!\!\!\!\!\!\!\!\!\!\!
{}_{*}^{*}\exp \left[ \sum_{n=2}^\infty \frac1{n!} \int \rd^4x_1\cdots
\rd^4 x_n \, W_n \Big(x_1,...,x_n;J[\vp]\Big)\,
\frac{\de}{\de \vp(x_1)} \cdots \frac{\de}{\de \vp(x_n)}\right]{}_{*}^{*}\;
\frac{\de S}{\de\vp(x)} \nn
\eea
Observe that for polynomial actions $S[\phi]$ the functional differential operator
reduces to a finite number of terms upon expansion of the exponential.

To have a concrete example, consider the classically conformal
$\phi^4$ theory with the action
\beq\label{S}
S[\phi] =\int\rd^4 x \left(\frac12 \partial_\mu \phi \partial^\mu \phi + \frac{\la}{4}\phi^4\right)
\eeq
This gives
\beq
\frac{\de\Ga[\vp]}{\de\vp(x)} =
\left\langle -\Box \phi(x) + \la \phi^3 (x) \right\rangle_{J=J[\vp]}
\eeq
and thus
\beq\label{DW}
\frac{\de\Ga[\vp]}{\de\vp(x)} =  - \Box\vp(x) + \la \vp^3(x) +
   3\la W_2(x,x;\vp) \vp(x) + \, \la W_3(x,x,x;\vp)
\eeq
Expressing $W_2$ and $W_3$ by means of (\ref{gga}) and (\ref{W3})
we see that all quantities in this equation can be expressed
in terms of $\Ga[\vp]$ and its functional derivatives, so that (\ref{DW})
indeed becomes a functional differential equation for $\Ga[\vp]$.

As they stand these equations, and in particular the basic functional equation
(\ref{Gvp}), are formal. Nevertheless,
there is already one useful application:  the equation (\ref{DW}) can be used as a
generating equation to derive the Schwinger-Dyson equations. With the
standard formula for the one-particle irreducible $n$-point functions
\beq
\Ga_n(x_1,\dots,x_n) \equiv \Ga_n(x_1,\dots, x_n;\vp) \Big|_{\vp = 0}
\eeq
we obtain, for instance,
\bea\label{Ga2}
\hbar\Ga_2(x,y) &=& \big( - \Box +  3\la W_2(x,x)\big) \de^{(4)} (x-y)  \nn\\[2mm]
&&\!\!\!\!\!\!\!\!\!\!\!\!\!\!\!\!
- \,  \la  \int \rd^4u\,\rd^4v\, \rd^4w \,W_2(x,u)W_2(x,v)W_2(x,w) \Ga_4(u,v,w,y)
\eea
which can be represented diagrammatically in the usual way. Similar
formulae for higher $n$-point can be deduced by repeated differentiation.

In principle, eq.~(\ref{DW}) is an {\em exact} non-linear functional
differential equation for the action functional $\Ga[\vp]$. In the full
{\em renormalized} theory, this functional should be well-defined on a
set of sufficiently well-behaved functions $\vp(x)$ (say, $C^\infty$
functions which fall of sufficiently rapidly at infinity). In addition,
its functional derivatives should be well-defined as distributions.
However, this cannot be the case for (\ref{DW}) as it stands. First of all,
the equation is written in terms of bare couplings and correlators, and
needs to be renormalized. Secondly, even if one assumes that the
necessary renormalizations have been performed, and the couplings are
replaced by the renormalized (physical) ones, (\ref{DW}) would still
not be well defined as it stands because the r.h.s. of (\ref{DW}) contains singular
contributions in the terms of order $\hbar$: recall that $G_2(x,y)$  and
higher $n$-point functions are generally singular at coincident points,
even in free field theory. It is for this reason that one conventionally must resort to perturbative methods by considering the $n$-point functions separately, and
by rendering them finite order by order in perturbation theory by means of
suitable subtractions in momentum space. For instance, this can be easily
seen from (\ref{Ga2}) where the infinity of $G_2(x,x)$ can be absorbed by an
appropriate wave function renormalization $\vp \rightarrow Z^{1/2} \vp$ at lowest order.

As already emphasized in the Introduction, we here adopt a different strategy 
by trying to deal with equation (\ref{DW}) {\em directly}. This requires to look for 
theories for which the DeWitt equation can be made well defined, that is, free 
of singularities. Examples of such theories are certain {\em supersymmetric models}
of the type discussed below in section~5. We note again that the DeWitt equation 
(\ref{Gvp}) is not of any known type. This is so even if one restricts this
equation to an `ordinary' partial differential equation for finitely many variables
as in the following section. This is one of the reasons for the difficulties in dealing with it, and motivates the present effort to gain a better understanding of this equation.


\section{A `zero-dimensional field theory' example}
To bring out  the main new features we now discuss an example from
`zero-dimensional' field theory, that is, a system with finitely many degrees of freedom,
in terms of which the results described in the foregoing section can be 
explicitly illustrated, and where we do not have to worry about UV infinities.
This example will also allow us to exhibit the vastly improved convergence properties
of a new summation scheme over conventional perturbation theory. To this aim
let us consider the `action' of a zero-dimensional $\phi^4$ theory
\beq
S(x) = \frac12 \sum_{i,j=1}^n x_i A_{ij} x_j + \frac{\la}4 \sum_{j=1}^n x_j^4
\label{findim}
\eeq
where $A_{ij}$ is a non-degenerate positive definite matrix. The generating function $W(J)\equiv W(J_1,\dots, J_n)$ for the `connected Green's functions' is then defined in analogy with (\ref{W}) as
\beq
e^{-W(J)} :=  
\int_{\mathbb{R}^n} \rd x\, \exp\big[ - S(x) - \sum_j x_j J_j \big]
\label{zerodimensionalint}
\eeq
where  the integration measure $\rd x$ is normalized in such a way that $W(0)=0$.
The generating function is easily seen to satisfy the differential equation
\beq
\sum_j A_{ij} \frac{\partial Z}{\partial J_j} + \la \frac{\partial^3 Z}{\partial J_i^3}= J_i Z(J)
\eeq
or, in terms of $W(J)$,
\beq
\sum_j A_{ij} \frac{\partial W}{\partial J_j}  +
\la\left[ \frac{\partial^3 W}{\partial J_i^3}
- 3\frac{\partial^2 W} {\partial J_i^2}\frac{\partial W}{\partial J_i}
 + \left( \frac{\partial W}{\partial J_i}\right)^3 \right] =  -J_i
\eeq
When expressed in terms of the effective action, this is the finite-dimensional
analog of the DeWitt equation (\ref{Gvp}), see below. So
in analogy with (\ref{vp})  let us define the `classical field' by
\beq
\vp_i(J) := \frac{\partial W(J)}{\partial J_i}
\eeq
and introduce the `effective action' $\Ga(\vp)$ in the usual way by Legendre
transformation as in (\ref{Gamma}). The DeWitt equation now reduces to a set of
partial differential equations
\beq\label{Gadiscrete}
\frac{\partial \Ga(\vp)}{\partial \vp_i} =
\exp\left[\sum_{k\geq 2} \frac1{k!}  \sum_{j_1,\dots,j_k}   W_{j_1\cdots\, j_k}(J)
\frac{\partial}{\partial \vp_{j_1}} \cdots \frac{\partial}{\partial \vp_{j_k}} \right]
\frac{\partial S(\vp)}{\partial \vp_i}
\eeq
where $W_{j_1\cdots j_k}\equiv (-1)^{k-1}\partial_{j_1} \cdots \partial_{j_k} W$, and we have
relations analogous to (\ref{gga}) and (\ref{W3}), that is, $\sum_j W_{ij} (J)
\Ga_{jk}(\vp(J)) = \delta_{ik}$, and so on.

We can now produce a formal solution of (\ref{Gadiscrete}), re-deriving a result
that was essentially obtained already long ago \cite{Jackiw}. From the general definition
we directly obtain the following differential equation for $\Ga(\vp)$
\bea\label{Ga}
&& \exp\left[ - \Ga(\vp) + \sum_j \vp_j \frac{\partial \Ga(\vp)}{\partial \vp_j}\right] = \nn\\[1mm]
&&  \qquad\qquad   = \; 
\int_{\mathbb{R}^n} \rd x\, \exp\left[ - S(x) +
            \sum_j x_j \frac{\partial \Ga(\vp)}{\partial \vp_j}\right]
\eea
To evaluate the integral we split the `effective action' into a `classical' part $S(\vp)$
and a `quantum' part $\tGa(\vp)$ according to
\beq
\Ga(\vp) =  \frac12 \sum_{i,j=1}^N \vp_i A_{ij} \vp_j + \frac14\la \sum_{j=1}^N \vp_j^4
      \, + \, \tGa(\vp)
\eeq
Shifting integration variables as $x_j \rightarrow x_j + \vp_j$ in (\ref{Ga}), a little algebra gives
\bea\label{tGa}
\exp\left[ - \tGa(\vp)\right] & = & \nn\\[1mm]
&& \!\!\!\!\!\!\!\!\!\!\!\!\!\!\!\!\!\!\!\!\!\!\!\!\!\!\!\!\!\!\!\!\!\!\!\!\!\!\!\!\!
 \int_{\mathbb{R}^n} \rd x\,
 \exp\left[ -\frac12 \sum_{ij} x_i G^{-1}_{ij}(\vp) x_j
    - \la\sum_j x_j^3 \vp_j -\frac{\la}4 \sum_j x_j^4 +
      \sum x_j \frac{\partial \tGa(\vp)}{\partial \vp_j} \right]   \nn\\[1mm]
\eea
with the classical `field-dependent' Green's function $G_{ij}(\vp)$
\beq\label{classProp}
\sum_j \big( A_{ij} + 3\la \delta_{ij} \vp_j^2 \big) G_{jk}(\vp) \,=\, \delta_{ik}
\eeq
Performing the Gaussian integral, and using Wick's theorem in the form
\bea\label{Wick}
(2\pi)^{-n/2} \int_{\mathbb{R}^n}  d^{n}x
f(x) \exp\left(-\frac{1}{2}\sum^{n}_{i,j=1}C_{ij}x_{i}x_{j}\right) &=& \nonumber\\[2mm]
&& \!\!\!\!\!\!\!\!\!\!\!\!\!\!\!\!\!\!\!\!\!\!\!\!\!\!\!\!\!\!\!\!
\!\!\!\!\!\!\!\!\!\!\!\!\!\!\!\!\!\!\!\!\!\!\!\!\!\!\!\!\!\!\!\!\!\!\!\!\!\!\!\!\!\!\!\!\!\!\!\!\!\!\!\!\!\!\!\!\!
=  (\det C )^{-1/2}\,
\exp\left(\frac{1}{2}\sum_{i,j=1}^n     C^{-1}_{ij}
\frac{\partial}{\partial y_i}\frac{\partial}{\partial y_j}\right) f(y)\bigg|_{y=0}
\eea
the expression (\ref{tGa}) can be re-written in the form
\bea\label{Gat}
\exp\left[ - \tGa(\vp)\right]
&=&  \big(\det G_{ij}(\vp)\big)^{1/2} \exp \left( \frac12 \sum_{i,j} G_{ij}(\vp)
\frac{\partial}{\partial \eta_i}\frac{\partial}{\partial \eta_j}\right) \nn\\[1mm]
&&  \quad  \exp\left[ - \la \sum_j \vp_j \eta_j^3  - \frac{\la}4 \sum_j \eta_j^4
         + \sum_j \eta_j \frac{\partial\tGa(\vp)}{\partial \vp_j} \right]_{\eta = 0}
\eea
Let us pause to explain this formula. The determinant prefactor just produces the
well known semi-classical (one-loop) correction $\propto \, \log \big(\det G_{ij}(\vp)\big)$ to the
classical action. As for the remaining terms, and ignoring the last term
$\propto \eta \partial\tGa/\partial\vp$, we would get the sum over all connected
vacuum diagrams with the field-dependent propagator $G_{ij}(\vp)$ 
(as the result of taking the logarithm on both sides). Although this
last term would seem to make the equation completely untractable, a little bit
of thought shows that this is not so.  Because $\tGa(\vp)$ contains only one-particle
irreducible contributions, the effect of this last term is precisely to remove the one-particle
reducible diagrams from the expansion: because this term is linear in $\eta$, it
can couple to the rest of any diagram only via a single line. In other words,
{\em the quantum effective action is nothing but the sum
of the one-loop correction and the sum over one-particle irreducible
vacuum diagrams with at least two loops and with the field-dependent
Green's function} (\ref{classProp}). This is the result derived in \cite{Jackiw}
for the effective potential in quantum field theory.

By construction, this series solution must satisfy the discrete DeWitt
equation (\ref{Gadiscrete}), and this claim can in principle be checked
order by order. Equally important is the fact that  the expansion, while being
asymptotic,  can have vastly better convergence properties for non-vanishing
$\vp$ than the usual perturbation expansion in terms of the coupling constant $\la$.
This is most easily seen by simplifying our zero-dimensional field theory even 
further to an integral over one variable. In this case the `Green's function' (\ref{classProp})
is simply $G(\vp) \equiv (1 + 3\la\vp^2)^{-1}$. For a given vacuum diagram
with $I$ internal lines we have
\beq
I=\frac32 V_3+2 V_4
\eeq
where $V_3$ and $V_4$, respectively, denote the number of three- and 
four-point vertices in (\ref{Gat}) ; note that in any vacuum diagram, the number
of three-point vertices is {\em even}.  The number of loops is equal to
\beq
L=\frac12 V_3+V_4+1
\eeq
Therefore an arbitrary vacuum diagram with $L$ loops will be proportional to
\beq\label{LoopOrder}
\frac{\la^{V_4}(\la\vp)^{V_3}}{(1+3\la\vp^2)^I}\approx
(\la\vp^4)^{1-L}
\eeq
(for $L=1$, the relevant parameter is $\log (1 + 3\la\vp^2)$).
In other words, the loop expansion now coincides with an expansion in $(\la\vp^4)^{-1}$:
of course, this expansion should only be used  in the appropriate region
in field space and the space of couplings, where $\la\vp^4$ is sufficiently large.
So we see that the series can converge well even for large $\la$ provided
the value of the classical field $\vp$ is not too small (and different from zero)!
We have checked this claim by numerical integration of a non-trivial example,
which we give in the Appendix. The important lesson, then, is that it is not simply 
the coupling constant $\la$ (or its running analog $\la(\mu)$, where $\mu$ is some renormalization scale) that governs the convergence properties of the effective 
action functional, but that one should also consider the question of convergence 
w.r.t. to the value of the field variables $\vp_j$ or $\vp(x)$ as well.

\section{Formal solution}
The considerations of the foregoing section can be straightforwardly extended
to field theory, enabling us to construct a formal expression for the (unrenormalized)
effective action {\em in terms of a sum over vacuum diagrams with field dependent
classical Green's functions}. For constant field configurations $\vp(x) = \vp_0$ this
solution reduces to the one found already long ago in  \cite{Jackiw}, where it
was exploited for an efficient determination of higher order corrections to the
Coleman-Weinberg effective potential for various theories. We here present the general
solution that allows for arbitrary $x$-dependence of the classical field $\vp$,
and that follows directly from the above construction by taking a formal limit
$n\rightarrow\infty$, or alternatively by a minor modification of the argument given
in \cite{Jackiw}. It is remarkable that in this way an explicit, albeit formal, solution of the
(unrenormalized) DeWitt equation can be obtained that would seem
difficult to guess otherwise. Of course, even in the full theory all relevant expressions 
can be made well defined by regulating the quantum field theory, either by discretization
as in the previous section, or by suitable continuum regularizations such
as smearing.

From (\ref{Gat}) we deduce immediately that the formal solution for the unrenormalized
effective action functional can be presented in the form
\bea\label{Jackiw}
\Ga[\vp] &=& S[\vp] \,+ \, \frac{\hbar}{2}\int d^{4}x
\log \left[\frac{\delta^{2}S[\vp]}{\delta\vp(x)\delta\vp(x)}\right] \\[2mm]
&&
\!\!\!\!\!\!\!\!\!\!\!\!\!\!\!\!\!\!\!\!\!\!\!\!\!\!
- \hbar \log \left[\exp \left(\frac{\hbar}2 \int d^4u d^4v \, G_{cl}(u,v;\vp)
\frac{\de^2}{\de\eta(u)\de\eta(v)}\right) \exp\left( -\hbar^{-1} S_{\rm int}[\vp,\eta]\right)\bigg|_{\eta =0}
       \right]_{\rm 1PI}  \nn
\eea
where the subscript 1PI means that one-particle reducible diagrams are to be omitted 
in the expansion, and where the logarithm  removes disconnected diagrams 
from inside the brackets. The interacting part  of the action
is defined by subtracting the linear and quadratic fluctuations,
\bea\label{Sint}
S_{\rm int}[\vp, \eta] &:=&  S[\vp + \eta] - S[\vp]
      -  \int d^4u \,\eta(u) \frac{\de S[\vp + \eta]}{\de \vp (u)}  \Bigg|_{\eta = 0} \nonumber\\
&&     \quad\quad   - \frac12 \, \int d^4u d^4  v \, \eta(u) \eta(v) \frac{\de^2 S[\vp + \eta]}{\de \vp(u) \de \vp(v)}
             \Bigg|_{\eta = 0} \nonumber\\
   &=&  \frac1{3!} \frac{\de^3 S}{\de \vp^3} \eta^3 +
             \frac1{4!} \frac{\de^4 S}{\de \vp^4} \eta^4 + \cdots
\eea
Observe that a residual dependence on $\vp$ arises from four-point vertices
onwards, whereas there is no $\vp$-dependence if there are only cubic vertices.
The expectation values in (\ref{Jackiw}) are to be computed with the classical
{\em field dependent} Green's function $G_{cl}(x,y;\vp)$, which is defined as
\beq
\int d^4y \, G_{cl}(x,y;\vp) \frac{\de^2S[\vp]}{\de\vp(y) \de\vp(z)}
  = \de^{(4)}(x-z)
\label{ggb}
\eeq
Hence $G_{cl}(x,y;\vp)$ is  the classical analog of (\ref{gga}),
in the sense that
\beq
W_2(x,y;\vp) = \hbar  G_{cl}(x,y;\vp)  + \cO(\hbar^2)
\eeq

According to the formula  (\ref{Jackiw}) the unrenormalized effective
action $\Gamma[\vp]$ is the {\em sum over all one-particle-irreducible (1PI) vacuum
diagrams with the field-dependent Green's function} (\ref{ggb}).  The dependence of $\Gamma$
on the field $\vp(x)$ thus derives from {\em two} sources, namely the field
dependence of $G_{cl}(x,y;\vp)$, and secondly the residual dependence of
$S_{\rm int}$ on $\vp$ (which only exists if there are 4-point or higher point vertices).
The former can be made more explicit by expanding
\beq
G_{cl}(x,y;\vp) = G_0(x,y) -  \int d^4u \, G_0(x,u) p(\vp(u)) G_0(u,y) \,\pm \, \cdots
\eeq
where $p(\vp)$ is obtained from $\de^2 S/\de\vp^2$ by removing the free part
not depending on $\vp(x)$, and $G_0$ is the free propagator.
The terms in this expansion thus  generate the `antenna-like' diagrams
known from textbook formulas of the effective potential.

By virtue of its definition and the above derivation, the expression
(\ref{Jackiw}) must satisfy the DeWitt equation (\ref{Gvp}) at least formally. This claim is straightforward to check for the semi-classical $\cO(\hbar)$ correction by use of the formula
\beq
\frac{\de}{\de\vp(x)} {\rm Tr} \log M = {\rm Tr} \left( M^{-1} \frac{\de M}{\de\vp(x)} \right)
\eeq
valid for any functional matrix $M(y,z)$, and by approximating the full two-point function $G(x,y;J(\vp))$ from (\ref{ggb}) by $G_{cl}(x,y;\vp)$. However, a direct verification of
(\ref{Jackiw}) to all orders is cumbersome. We will therefore postpone a discussion
of this issue to the following section in terms of an example where the
DeWitt equation is well defined. Let us just note that
in conjunction with the explicit expression as a sum over $\vp(x)$-dependent
vacuum diagrams we can see directly from the DeWitt equation (\ref{Gvp})
that $\Gamma[\vp]$ can only contain one-particle irreducible (1PI) diagrams:
the action of the first functional derivative $\delta\Gamma[\vp]/\delta\vp(x)$
in particular leads to the cutting any one of the propagators in a diagram
arising in the expansion (\ref{Jackiw}). If we had a diagram which is not
1PI then there would be at least one propagator which joins two 1PI subdiagrams.
The action of the functional derivative on this diagram would thus split the diagram
in two parts at this propagator, leaving two disconnected diagrams. But on the r.h.s.
of the DeWitt equation we have only connected Green's functions,
$\delta^{n}W[J]/\delta J(x_1)\cdots\delta J(x_n)$ . So there can be no
disconnected diagrams on the r.h.s. of \cite{Jackiw} and thus we can only have
1PI diagrams contributing to $\Gamma[\vp]$, as expected.

The effective (Coleman-Weinberg) potential is obtained by specializing all
formulas to $x$-independent fields $\vp(x)=\vp_0$ \cite{CW} and removing a
formally infinite volume factor $\propto \int dx$.  The main
advantage of writing the effective potential as a sum over vacuum type
diagrams is the following: rather than having to do all the combinatorics with
`antenna diagrams', one obtains the answer at each loop order `in one stroke'.
In particular the RG improved one-loop potential obtained by summing ladder
bubble diagrams is directly obtained. This was, in fact, the first application
of this formula in \cite{Jackiw} where the effective potential as also determined to
two loops for $\vp^4$ theory. As shown there the formalism implies
considerable simplifications in comparison with the textbook derivations
of the Coleman-Weinberg potential.

At the end of this section we write the solution (\ref{Jackiw}) for the 
finite dimensional integral with the action defined by (\ref{findim}), that is,
the  solutions to (\ref{Ga}). In accordance with the explanation after
(\ref{Jackiw}) we include only 1PI and connected diagrams 
in the expansion
\beq
\Ga(\vp_i)=S(\vp_i)+\Ga^{(1)}(\vp_i)+\Ga^{(2)}(\vp_i)+\Ga^{(3)}(\vp_i)+\ldots
\label{effactionexplicit}
\eeq
where the indices denote the loop order. In this way we obtain
\bea
\Ga^{(1)}(\vp) &=& - \, \frac{1}{2}
\ln \det(G_{ij}) \nn\\
\Ga^{(2)}(\vp) &=&
- \left[-\frac{3\la}{4}\sum_i G_{ii}^2+3\la^2\sum_{i,j}\vp_i\vp_j G_{ij}^3\right]  \nn\\
\Ga^{(3)} (\vp)&=&
-\left[\frac{3\la^2}{4}\sum_{i,j}G_{ij}^4+
\frac{9\la^2}{4}\sum_{ij}G_{ii}G_{ij}^2G_{jj}\right.\nn\\
&&
-27\la^3\sum_{i,j,k}\vp_i\vp_jG_{ij}G_{ik}^2G_{jk}^2
-27\la^3\sum_{i,j,k}\vp_i\vp_jG_{ij}^2G_{ik}G_{jk}G_{kk}\nn\\
&&+54\la^4\sum_{i,j,k,l}\vp_i\vp_j\vp_k\vp_l
G_{ij}G_{jk}G_{kl}G_{li}G_{ik}G_{jl}\nn\\
&&\left.
+81\la^4\sum_{i,j,k,l}\vp_i\vp_j\vp_k\vp_l
G_{ij}^2G_{kl}^2G_{ik}G_{jl}\right]\nn
\eea
As already pointed out, this is a `nonperturbative expansion' that is restricted 
to the region of couplings and field space where the `parameter' 
$G(\vp) \sim (\la\vp^2)^{-1}$ is small. In the formula above we included terms 
up to three loops, i.e. up to sixth order in $G_{ij}(\vp)$ (one easily checks that
all terms are of the appropriate order in $(\la\vp^4)^{-1}$, in agreement with formula
(\ref{LoopOrder})). A numerical comparison of the exact 
result and this expansion for a one-dimensional integral for several 
values of $\la$ and $\vp$ is given in the appendix. It shows that this expansion
can give excellent agreement with the exact result even in regions where
$\la$ is very large.

\section{The Wess-Zumino model in $D=2$}

We next turn to an example where the DeWitt equation (\ref{Gvp}) can be made completely well-defined, that is, free of all short distance singularities. This is the $N=1$ Wess-Zumino model in two space-time dimensions which is UV finite order by order in perturbation theory (the generic non-supersymmetric theories having only logarithmic divergences in two dimensions, which are removed by imposing supersymmetry).~\footnote{See
   \cite{Maluf} for a recent treatment of the Wess-Zumino model in 2+1 dimensions.}

The Euclidean version of the model can be written in terms of a single superfield
$\Phi(z)$ with superspace coordinate $z\equiv (x,\theta)$, where $\theta$ is a
two-component (anti-commuting) Majorana spinor with $\theta=\theta^{*}$. The
superfield contains a real scalar $A$ and a Majorana spinor $\psi$, as well
as the auxiliary field $F$:
\beq
 \Phi (x,\theta) =A(x)+\bar{\theta}\psi(x)+\frac{1}{2}\bar{\theta}\theta F(x)\\
\eeq
For simplicity we restrict attention to the following Lagrangian
\beq
\cL= -\frac{1}{4}\Phi\bar{D}D\Phi+\frac{1}{2}m\Phi^{2}+\frac{1}{3}g\Phi^{3}
\eeq
We could replace the last two terms by an arbitrary polynomial $P(\Phi)$ here,
but this would only make the formulas more cumbersome and not give any
new insights. The supercovariant derivative is defined by
\beq
 D_{\alpha}=\frac{\partial}{\partial\bar{\theta}^{\alpha}}
 +(\gamma^{\mu}\theta)_{\alpha}\partial_{\mu}
\eeq

\beq
\bar{D}^{\alpha}=-C^{\alpha\beta}D_{\beta}
\eeq
where $C$ is the charge conjugation matrix.
The lagrangian in component form is as follows:
\beq
\cL = \frac{1}{2}A\Box A-\frac{1}{2}\bar{\psi}\gamma^{\mu}\partial_{\mu}\psi+\frac{1}{2}
F^{2}+\frac{1}{2}m(2AF-\bar{\psi}\psi)+g(A^{2}F-A\bar{\psi}\psi)
\eeq
Writing out the DeWitt equation for the three fields $A,\psi$ and $F$ we get
\bea
\frac{\delta\Gamma[A,F,\psi]}{\delta A(x)} &=& \Box A(x)+mF(x)+g\big[ 2A(x)F(x)-\bar{\psi(x)}\psi(x)\big]\nn\\
&& \quad
- \, g \hbar\left[ 2\, \frac{\delta^{2}W[J]}{\delta J_A (x)\delta J_F(x) }
+ \, {\rm Tr}\, \frac{\de^2 W[J]}{\de J_\psi(x) \de J_{\bar\psi}(x)} \right]\nn\\[2mm]
\frac{\delta\Gamma[A,F,\psi]}{\delta\bar{\psi}(x)} &=&-\slashed{\partial}\psi(x)-m\psi(x)-2gA(x)\psi(x)-\hbar g \frac{\delta^{2}W[J]}{\delta J_{A}(x)\delta J_{\psi}(x)} \nn\\[2mm]
\frac{\delta\Gamma [A,F,\psi]}{\delta F(x)}&=&
   F(x)+mA(x)+gA^{2}(x)-\hbar g \frac{\delta^{2}W[J]}{\delta J_{A}(x)\delta J_{A}(x)}
\eea
with self-explanatory notation.
Now we see that the equation for the scalar field $A$ is well defined as it stands
because the logarithmic singularities cancel between the two terms in parentheses.
More precisely, the latter expression is understood to be
\beq
\lim_{y\rightarrow x} \left[ 2\, \frac{\delta^{2}W[J]}{\delta J_A (x)\delta J_F(y) }
+ \, {\rm Tr}\, \frac{\de^2 W[J]}{\de J_\psi(x) \de J_{\bar\psi}(y)} \right]
 = \mbox{finite}
\eeq
Likewise the equation for $\psi$ is well defined because $\de^2 W/\de A\de \psi$ is
free of short distance singularities. So the only singularity occurs in the last equation,
and this can be removed by replacing the product $A^2(x)$ by the normal ordered product
\beq
:\! A(x)A(y) \! :  \;  \equiv A(x) A(y)  -  \underbracket{A(x)A(y)}
\eeq
and taking $x\rightarrow y$ afterwards. This singularity simply  follows from the
fact that if one expresses the auxiliary field $F$ in terms of the physical field $A$, the non-linear terms in $A$ must be rendered non-singular to make $F$ itself well-defined as a quantum operator. \footnote{But note that, while $:\!A^2\!:$ is well-defined as an operator,
   it is singular as a $c$-number, while the converse is true for $A^2$!}
Consequently, the last component of the DeWitt equation must be replaced by
\beq
\frac{\delta\Gamma [A,F,\psi]}{\delta F(x)} =
   F(x)+mA(x)+g:\! A^{2}(x)\!: -\hbar g \frac{\delta^{2}W[J]}{\delta J_{A}(x)\delta J_{A}(x)}
\eeq
and then {\em all} components of the DeWitt equation are free of singularities.
In practice, the above replacement simply means that in the formal solution as
a sum over vacuum diagrams there are no tadpole diagrams (these are anyway
absent for a theory with only cubic vertices as they would lead to
non-1PI diagrams in $\Ga$ which cannot be).

All these equations can be conveniently recast into superspace equations.
A similar normal ordering can be done in the superspace version of the lagrangian and as it is much more convenient to work in it we would stick to the superspace description.
So we have the functional derivative of the action as:
\beq
\frac{\delta S}{\delta{\Phi}}= -\frac{1}{2}\bar{D}D\Phi+m\Phi+g\Phi^{2}
\eeq
The arguments of the foregoing sections generalize directly to superspace. For the cubic Lagrangian above the DeWitt equation (\ref{Gvp}) takes an especially simple form, namely
\beq\label{superDW}
\frac{\de\Ga[\Phi]}{\de\Phi(z)} = \; : \!\frac{\de S[\Phi]}{\de\Phi(z)} \!: \: -\:
\hbar g \frac{\de^2 W[J]}{\de J(z)\de J(z)} \bigg|_{J=J[\Phi]}
\eeq
or, more specifically
\beq\label{superDW1}
\frac{\delta\Gamma[\Phi]}{\delta \Phi(z)}=-\frac{1}{2}\bar{D}D\Phi(z) +m\Phi(z)+g:\! \Phi^{2}(z)\! :-
\hbar g \frac{\de^2 W[J]}{\de J(z)\de J(z)} \bigg|_{J=J[\Phi]}
\eeq
where $z\equiv (x^\mu, \theta)$ and $J(z)$ is the `supersource field'
$J(z) \equiv J_F + \bar\theta J_\psi + \bar\theta \theta J_A$. The normal ordering is
understood to be in the sense of the component expressions given above.
In the formal solution below this simply means that all tadpole diagrams
are suppressed.

For the free superfield the superspace propagator is
\bea
&&\!\!\!\!\!\!\!\!\!\!\!\!\!\!\!\!
G_{2}^{(0)}(z - z')\, = \nn\\
&&\!\!\!\!\!\!\!\!\!\!\!\!\!\!\!\!
=\big\langle 0\big\rvert T[(A(x)+\bar{\theta}\psi(x)+\frac{1}{2}\bar{\theta}\theta F(x))(A(x')+\bar{\theta}'\psi(x')+\frac{1}{2}\bar{\theta}'\theta' F(x'))]\big\lvert 0\big\rangle\nn\\
&&\!\!\!\!\!\!\!\!\!\!\!\!\!\!\!\!
=\exp \Big[ -\frac{1}{2}(\bar{\theta}-\bar{\theta}')(\gamma_{\mu}\partial^{\mu}+m)
(\theta-\theta')\Big]\triangle_{F}(x-y)
\eea
In analogy with (\ref{ggb}) we define the Green's function in superspace
\beq
\int dz' \, G_{cl}(z,z';\Phi)\, \frac{\de^2S[\Phi]}{\de\Phi(z') \de\Phi(z'')}
  = \de (z-z'')
\label{ggb1}
\eeq
(where the fermionic part of the $\de$-function is defined in the usual way as
$\de(\theta) = \theta$) so that $G_{cl}(z,z';\Phi) = G_2^{(0)}(z-z') + \cdots$.

By construction the supersymmetric DeWitt equation (\ref{superDW}) is well defined,
and we can therefore take over the formal solution given in the previous section,
\bea
\Gamma[\Phi] &=&
S[\Phi] \,+\, \frac{\hbar}{2}\int d^{4}z \ln \left[\frac{\delta^{2}S}{\delta\Phi(z)\delta\Phi(z)}\right]
\nn\\[2mm]
&& -\, \hbar \ln\bigg[\exp\left( \frac{\hbar}{2}G_{i,j}\frac{\delta^{2}}{\delta\tilde{\Phi_{i}}\delta\tilde{\Phi_{j}}}\right)
\exp\left( -\frac{\tilde{S}_{int}}{\hbar}\right) \bigg|_{\tilde{\Phi}=0}\bigg]
\eea
Where $G_{ij}$ is shorthand for  $G_{cl}(z_{i},z_{j};\Phi)$ and
$\tilde{S}_{int}=\frac{g}{3}\tilde{\Phi}^{3}$, and all the integrals are
understood to be in superspace. Now if we expand the series we have the following:
\beq
\bigg[1+\sum_{n=1}^{\infty}\frac{1}{n!}\bigg(\frac{\hbar}{2}G_{i,j}
\frac{\delta^{2}}{\delta\tilde{\Phi_{i}}\delta\tilde{\Phi_{j}}}
\bigg)^{n}\bigg]\bigg[1+\sum_{m=1}^{\infty}\frac{1}{m!}
\bigg(\frac{-\tilde{S}_{int}}{\hbar}\bigg)^{m}\bigg]\bigg|_{\tilde{\Phi}=0}
\eeq
Because the dummy variable $\tilde{\Phi}$ is put to 0, and the interaction is cubic,
only terms with $2n=3m$ survive. Thus the first of this will be at two loops for $m=2, n=3$.
 Evaluating the corresponding term we get
\beq
\bigg[\frac{1}{3!}\bigg(\frac{\hbar}{2}G_{i,j}
\frac{\delta^{2}}{\delta\tilde{\Phi_{i}}\delta\tilde{\Phi_{j}}}\bigg)^{3}\bigg]
\bigg[\frac{1}{2!}\bigg(\frac{-\tilde{S}_{int}}{\hbar}\bigg)^{2}\bigg]\nn\\
=\frac{\hbar g^{2}}{3}\int_{z,w} G^{3}_{cl}(z,w;\Phi)\nn
\eeq
At the next order (three loops) we have $n=6, m=4$, and
\bea
&&\!\!\!\!\!\!\!\!\!\!\!\!\!\!\!\!
\bigg[\frac{1}{6!}\bigg(\frac{\hbar}{2}G_{i,j}
\frac{\delta^{2}}{\delta\tilde{\Phi_{i}}\delta\tilde{\Phi_{j}}}\bigg)^{6}\bigg]
\bigg[\frac{1}{4!}\bigg(\frac{-\tilde{S}_{int}}{\hbar}\bigg)^{4}\bigg]\nn\\[3mm]
&&\!\!\!\!\!\!\!\!\!\!\!\!\!\!\!\!
=\hbar^{2}\bigg[\frac{2}{3}g^{4}
\int_{u,v,w,z}G_{cl}(u,v;\Phi)G_{cl}(u,w;\Phi)G_{cl}(u,z;\Phi)
G_{cl}(v,w;\Phi)G_{cl}(v,z;\Phi)G_{cl}(w,z;\Phi)\nn\\[2mm]
&&\!\!\!\!\!\!\!\!\!\!\!\!\!\!\!\!
+g^{4}\int_{u,v,w,z}G^{2}_{cl}(u,v;\Phi)G^{2}_{cl}(w,z;\Phi)G_{cl}(u,w;\Phi)G_{cl}(v,z;\Phi)+
\frac{1}{2}\bigg(\frac{g^{2}}{3}\int_{z,w} G^{3}_{cl}(z,w;\Phi)\bigg)^{2}\bigg]\nn\\
\eea
We recognize the last term as square of the term which we got for $n=3,m=2$ (two loops), which is removed by taking log of the entire expression as these diagrams are not connected. Hence summing up we get the following contribution to the effective action:
\bea
&&\!\!\!\!\!\!\!\!\!\!\!\!\!\!\!\!
\Gamma=S+\frac{\hbar}{2}\int d^{4}z
\ln \left[\frac{\delta^{2}S}{\delta\Phi(z)\delta\Phi(z)}\right]
-\frac{\hbar^{2}g^{2}}{3}\int_{z,w} G_{cl}^{3}(z,w;\Phi)\nn\\[2mm]
&&\!\!\!\!\!\!\!\!\!\!\!\!\!\!\!\!
-\frac{2}{3}\hbar^{3}g^{4}\int_{u,v,w,z}G_{cl}(u,v;\Phi)G_{cl}(u,w;\Phi)
G_{cl}(u,z;\Phi)G_{cl} (v,w;\Phi)G_{cl} (v,z;\Phi) G_{cl}(w,z;\Phi)\nn\\[2mm]
&&\!\!\!\!\!\!\!\!\!\!\!\!\!\!\!\!
-\hbar^{3}g^{4}\int_{u,v,w,z}G_{cl}^{2}(u,v;\Phi)G_{cl}^{2}(w,z;\Phi)
G_{cl}(u,w;\Phi)G_{cl}(v,z;\Phi)      \, + \, \cO(\hbar^4)
\eea
To check this we first calculate the second functional derivative of $\Ga$, which is,
up to order $\hbar^{2}$,
\bea
&&\!\!\!\!\!\!\!\!\!\!\!\!\!\!\!\!
\frac{\delta^{2}\Gamma}{\delta\Phi(z_{1})\delta\Phi (z_{2})}=\frac{\delta^{2}S}{\delta\Phi(z_{1})\delta\Phi(z_{2})}-2\hbar g^{2}G_{cl}(z_{1},z_{2};\Phi)G_{cl}(z_{1},z_{2};\Phi)\nn\\[2mm]
&&\!\!\!\!\!\!\!\!\!\!\!\!\!\!\!\!
-8\hbar^{2}g^{4}\int_{z,w} G_{cl}(z,z_{2};\Phi)G_{cl}(z_{2},z_{1};\Phi)
G_{cl}(z_{1},w;\Phi)G_{cl}^{2}(z,w;\Phi)\nn\\[2mm]
&&\!\!\!\!\!\!\!\!\!\!\!\!\!\!\!\!
-8\hbar^{2}g^{4}\int_{z,w} G_{cl}(z,z_{1};\Phi)G_{cl}(z_{1},w;\Phi)
G_{cl}(z,z_{2};\Phi)G_{cl}(z_{2},w;\Phi)G_{cl}(z,w;\Phi)\nn\\
\eea
Inverting the above we obtain the 2-point function up to order $\hbar^{2}$,
\bea
&&\!\!\!\!\!\!\!\!\!\!\!\!\!\!\!\!
-\frac{\delta^{2}W}{\delta J(z_{1})\delta J(z_{2})}=G_{cl}(z_{1},z_{2};\Phi)+2\hbar g^{2}\int_{u,v}G_{cl}(z_{1},u;\Phi)G_{cl}^{2}(u,v;\Phi)G_{cl}(v,z_{2};\Phi)\nn\\[2mm]
&&\!\!\!\!\!\!\!\!\!\!\!\!\!\!\!\!
+8h^{2}g^{4}\int_{u,v,z,w}\bigg[G_{cl}(z_{1},u;\Phi)G_{cl}(z,v;\Phi)
G_{cl}(v,u;\Phi)G_{cl}(u,w;\Phi)G_{cl}^{2}(z,w;\Phi)G_{cl}(v,z_{2};\Phi)\nn\\[2mm]
&&\!\!\!\!\!\!\!\!\!\!\!\!\!\!\!\!
+G_{cl}(z_{1},u;\Phi)G_{cl}(z,u;\Phi)G_{cl}(u,w;\Phi)
G_{cl}(z,v;\Phi)G_{cl}(v,w;\Phi)G_{cl}(z,w;\Phi)G_{cl}(v,z_{2};\Phi)\bigg]\nn\\[2mm]
&&\!\!\!\!\!\!\!\!\!\!\!\!\!\!\!\!
+4\hbar^{2}g^{4}\int_{u,v,w,z}G_{cl}(z_{1},u;\Phi)G_{cl}^{2}(u,v;\Phi)
G_{cl}(v,w;\Phi)G_{cl}^{2}(w,z;\Phi)G_{cl}(z,z_{2};\Phi)
\eea
Now putting this in the DeWitt equation from the R.H.S, we obtain:
\bea
&&\!\!\!\!\!\!\!\!\!\!\!\!\!\!\!\!
\frac{\delta S}{\delta \Phi(z)} -\hbar g \frac{\delta^{2}W}{\delta J(z)\delta J(z)}\nn\\[3mm]
&&\!\!\!\!\!\!\!\!\!\!\!\!\!\!\!\!
=\frac{\delta S}{\delta \Phi(z)}+\hbar gG_{cl}(z,z;\Phi)+
2\hbar^{2} g^{3}\int_{u,v}G_{cl}(z,u;\Phi)G_{cl}^{2}(u,v;\Phi)G_{cl}(v,z;\Phi)\nn\\[2mm]
&&\!\!\!\!\!\!\!\!\!\!\!\!\!\!\!\!
+8h^{3}g^{5}\int_{u,v,z',w}\bigg[G_{cl}(z,u;\Phi)G_{cl}(z',v;\Phi)
G_{cl}(v,u;\Phi)G_{cl}(u,w;\Phi)G_{cl}^{2}(z',w;\Phi)G_{cl}(v,z;\Phi)\nn\\[2mm]
&&\!\!\!\!\!\!\!\!\!\!\!\!\!\!\!\!
+G_{cl}(z,u;\Phi)G_{cl}(z',u;\Phi)G_{cl}(u,w;\Phi)
G_{cl}(z',v;\Phi)G_{cl}(v,w;\Phi)G_{cl}(z',w;\Phi)G_{cl}(v,z;\Phi)\bigg]\nn\\
&&\!\!\!\!\!\!\!\!\!\!\!\!\!\!\!\!
+4\hbar^{3}g^{5}\int_{u,v,w,z'}G_{cl}(z,u;\Phi)
G_{cl}^{2}(u,v;\Phi)G_{cl}(v,w;\Phi)G_{cl}^{2}(w,z';\Phi)G_{cl}(z',z;\Phi)
\eea
This is exactly what we get from the l.h.s. by taking the
first functional derivative of $\Ga$.

\section{Liouville Field Theory}

As an example where the DeWitt equation can be worked out explicitly
for  a theory with a {\em non-polynomial} action, we briefly consider Liouville theory
in two dimensions. As is well known, the actual construction of this special
conformal theory is subtle and has a long history (see e.g. \cite{dorn,zz} and 
references therein), so we here content ourselves with formal arguments and derivations, 
postponing a more detailed discussion to future work. We note that the derivations 
given below do not make any use of the conformal symmetry of the theory.

The generating functional $W[J]$ is defined as in (\ref{W}) with the action
\beq\label{SLiouville}
S=\int d^{2}x \left[\frac{1}{2}(\partial_{\mu}\phi)^{2}+\mu e^{b\phi(x)}\right]
\eeq
where we set $\hbar =1$ for simplicity. The proper definition of the theory is tricky, not
least because the run-away nature of the exponential potential does
not allow for proper classical vacuum. As a consequence,
the one-point function, and thus the classical field $\vp(J)$ 
may not be well defined  in all circumstances for this reason: in fact,
we would expect it to exist only for sources obeying $J(x) < 0$, for which the
potential valley is avoided.

For Liouville theory the main interest is not with expectation values
of products of field operators $\phi(x)$, but rather with the expectation values
of proper {\em primary fields}, which are exponential operators of the form
\beq
V_\alpha(x) \equiv \exp\big(\alpha\phi(x)\big)
\eeq
The correlation functions are then given by
\beq
\langle 0\rvert V_{\alpha_{1}}(x_{1})\cdots V_{\alpha_{n}}(x_{n})\lvert 0 \rangle =
\int \cD\phi \, e^{\alpha_{1}\phi(x_{1})} \cdots e^{\alpha_{n}\phi(x_{n})}e^{-S[\phi]}
\eeq
with the action (\ref{SLiouville}) (where again we assume proper normalization of the
path integral). Introducing a source $J(x)$ as before,
the correlation functions can be represented by means of a
$J$-dependent partition function, with a distributional source
\beq\label{Jn}
J(x) = -\sum_{j=1}^n \alpha_j \delta (x-x_j)
\end{equation}
The DeWitt equation can be worked out as before, with the result
\bea
&&\!\!\!\!\!\!\!\!\!\!\!\!
\frac{\delta\Gamma[\vp]}{\delta\vp(x)}= \\[2mm]
&&\!\!\!\!\!\!\!\!\!\!\!\!
{}_{*}^{*}\exp\bigg[\sum^{\infty}_{n=2}\frac{1}{n!}\int d^{4}x_{1}....d^{4}x_{n}W_n(x_{1},...,x_{n};J[\vp])
\frac{\delta}{\delta\vp(x_{1})}\cdots \frac{\delta}{\delta\vp(x_{n})}\bigg]{}_{*}^{*}\nn\\
&&\bigg[-\Box \vp(x)
+\mu be^{b\vp(x)}\bigg]   \nn
\eea
and thus
\bea
&&\!\!\!\!\!\!\!\!\!\!\!\!\!\!\!\!
\frac{\delta\Gamma[\vp]}{\delta\vp(x)}=-J(x)=-\Box\phi(x)+
\exp\bigg[\sum^{\infty}_{n=2}\frac{b^n}{n!}W_{n}(x,...,x;J[\vp])\bigg]\mu be^{b\vp(x)}\nn\\[2mm]
&&\!\!\!\!\!\!\!\!\!\!\!\!\!\!\!\!
\Rightarrow\log\big(-J(x)+\Box\vp(x)\big)
=\bigg[\sum^{\infty}_{n=2}\frac{b^n}{n!}W_{n}(x,...,x;J[\vp]) \bigg]+\ln(\mu b) +b\vp(x)\nn\\
&&\!\!\!\!\!\!\!\!\!\!\!\!\!\!\!\!
\eea
From (\ref{Wn0})  we know that,
\beq
W_{n}(x_{1},\dots ,x_{n})=(-1)^{n-1}\frac{\delta^{n}W[J]}{\delta J(x_{1})\cdots\delta J(x_{n})}\nn\\
\eeq
After some algebra, we obtain:
\bea\label{DWL}
\log \big(-J(x)+\Box\phi(x)\big) &=&
- \exp \left( - b \frac{\delta}{\delta J(x)}\right) W[J]
+\ln(\mu b)+W[J]\nn\\[2mm]
 &=& -W[J-b\delta_{x}]+W[J]+\ln(\mu b)
 \eea
 with $\delta_x(y) \equiv \delta (x-y)$. Equivalently, we can write
\beq
 -J(x)+\Box\phi(x)=\mu b e^{-W[J-b\delta_{x}]}e^{+W[J]}
      \equiv b\mu\frac{Z[J-b\delta_{x}]}{Z[J]}\nn\\
\eeq
Thus,  for $J$ of the form (\ref{Jn}), $Z[J-b\delta_{x}]$ has one more 
insertion than $J$, so the ratio appearing on the 
right-hand side in the previous equation is just
\beq
\frac{\langle 0\rvert V_{\alpha_{1}}(x_{1})\cdots V_{\alpha_{n}}(x_{n})
         V_{b}(x)\lvert 0 \rangle}{\langle 0\rvert V_{\alpha_{1}}(x_{1})\cdots 
         V_{\alpha_{n}}(x_{n})\lvert 0 \rangle}
\eeq
The usefulness of the equation (\ref{DWL})  is still under study, and we intend to
return to it in future work. At this point we only remark that,
if we define $\vp(x)$ as
\beq
\frac{\partial_{\alpha}\int [\textit{D}\phi]e^{\alpha\phi(x)}e^{-S[\phi]-J\cdot\phi}\lvert_{\alpha=0}}{\int [\textit{D}\phi] \,e^{-S[\phi]-J\cdot \phi}}
\eeq
the equation can be rewritten as follows:
\bea
&& (-J(x)) \langle 0\rvert\prod_{i=1}^{n}V_{\alpha_{i}}(x_{i})\lvert 0\rangle+\Box\bigg(\partial_{\alpha}         \langle 0\rvert V_{\alpha}(x)\prod_{i=1}^{n}
V_{\alpha_{i}}(x_{i})\lvert 0\rangle\lvert\bigg)_{\alpha = 0}\nn\\[2mm]
&& \qquad\qquad =b\mu\langle 0\rvert V_{b}(x)\prod_{i=1}^{n}V_{\alpha_{i}}(x_{i})\lvert 0\rangle
\eea
A similar equation, minus the first term was used by \cite{dorn}, to check the proposal
for the three point function in Liouville theory. If we plug in the DOZZ proposal in this
equation then we find that, neglecting contact terms,
\beq
4(\Delta_{1}-\Delta_{2})^{2}\partial_{\alpha}C(\alpha,\alpha_{1},\alpha_{2})|_{\alpha=0}=b\mu C(b,\alpha_{1},\alpha_{2})
\eeq
where the $\triangle_{i}$ are the conformal dimensions of the primary operators.
The other term which arises from the contact term, i.e. the $\delta$-functions obtained
by the action of the Laplacian, cancels with term proportional to $J$, generating
the on-shell constraint, $\alpha= \frac12 Q$.

\section{Outlook}

In the Introduction we already mentioned possible further directions. In particular, 
we would like to apply the DeWitt equation to $N=4$ Yang Mills  theory,
the prime example of a UV finite quantum field theory in four space-time
dimensions. However, this is not as straightforward as one might have wished.
One main obstacle is the lack of a fully off-shell supersymmetric realization 
of the theory. If we simply use the on-shell supersymmetric formulation (in Wess-Zumino
gauge), there will appear all kinds of spurious divergences, since only gauge invariant
observables are supposed to be UV finite. The same trouble would arise with
formulations where only part of the supersymmetry is realized off-shell (for instance,
in a formulation of the theory in terms of $N=1$ superfields), or with harmonic 
superspace. One could also try the opposite approach, where only the true on-shell
degrees of freedom are used, namely the light-cone superspace formalism 
proposed in \cite{Brink1}. There, the lagrangian is written in terms of a single 
chiral superfield using only physical degrees of freedom of the theory; using 
Grassmann parameters $\theta^{m}$ and their complex conjugates $\bar{\theta}_{m}$. 
The lagrangian for $N=4$ Yang Mills theory then takes the
following form \cite{Brink,Brink1}
\bea
L=&72\Big[-\bar{\phi}^{a}(\frac{\Box}{\partial^{+2}})\phi^{a}+
\frac{4}{3}gf^{abc}[\frac{1}{\partial^{+}}\bar{\phi}^{a}\phi^{b}\bar{\partial}\phi^{c}+
\frac{1}{\partial^{+}}\phi^{a}\bar{\phi}^{b}\partial\bar{\phi}^{c}]\notag\\[2mm]
&-g^{2}f^{abc}f^{ade}[\frac{1}{\partial^{+}}(\phi^{b}\partial^{+}\phi^{c})\frac{1}{\partial^{+}}
(\bar{\phi}^{d}\partial^{+}\bar{\phi}^{e})+
\frac{1}{2}\phi^{b}\bar{\phi}^{c}\phi^{d}\bar{\phi}^{e}]\Big]\notag\\
\eea
Using this Lagrangian we can formally write down a well defined DeWitt equation 
for this model. However, we have found that the resulting expressions are rather
messy, mainly because one has to keep track of all the non-local
$\partial_{+}^{-1}$ operator insertions. A more promising avenue seems to be 
that one should try to link up with very recent advances in the computation
of gauge theory and supersymmetric Yang-Mills amplitudes \cite{Bern1,Bern2}.
Although this formalism is {\em on-shell}, whereas the effective action functional is
by definition off-shell, very recent work \cite{Staudacher} indicates that it might
be possible to arrive at a formulation which is not off-shell in the momenta
$p_{\alpha\dot\beta} \equiv p_\mu \sigma^\mu_{\alpha\dot\beta}$,
but would be off-shell in the twistor-like variables $\chi_\alpha$ and
$\tilde\chi_{\dot\beta}$ used to represent on-shell momenta via
$p_{\alpha \dot\beta} = \chi_\alpha\tilde\chi_{\dot\beta}$. Clearly, this would
lead to an entirely new formulation of quantum field theory and the effective action.

\vspace{4.0mm}
\noindent
{\bf Acknowledgments:}  H.N. is grateful to H. Dorn, D. Kreimer, G. Jorjadze 
and M. Staudacher for discussions. K.A.M. thanks the Albert-Einstein-Institut for
hospitality and support. The work of Parikshit Dutta is supported by the 
Erasmus Mundus Joint Doctorate Program by Grant Number 2010-1816 from the EACEA of the European Commission.


\section{Appendix: Numerical results.}
To illustrate the efficiency of the expansion (\ref{Jackiw}) we 
present some numerical results for the simple one-dimensional integral
\beq
\exp(-W(J)) = \int \frac{dx}{\sqrt{2\pi}} \exp\left[ -\frac12 x^2 - \frac{\la}4 x^4 - xJ\right]
\eeq
in this appendix. To this aim, we go through the same steps
as before, with the expansion (\ref{effactionexplicit}) 
and $n=1$ in (\ref{zerodimensionalint}). The loop expansion
(\ref{GaLoop}) here becomes
\bea
\Sigma^{(0)}(\vp) &\equiv& S_{\rm cl} (\vp) =\frac{\vp^2}{2}+\frac{\la\vp^4}{4}\nn\\
\Sigma^{(1)}(\vp) &=& S_{\rm cl}-\frac12\ln(G)\nn\\
\Sigma^{(2)}(\vp) &=&\Sigma^{(1)}- \left(-\frac{3\la}{4}G^2+3\la^2\vp^2G^3\right) \nn \\
\Sigma^{(3)} (\vp) &=&\Sigma^{(2)}-\left(\frac{3\la^2}{4}G^4+\frac{9\la^2}{4}G^4
-27\la^3\vp^2 G^5-27\la^3\vp^2G^5\right.\nn\\
&&\left.\ \ \ \ \ \ \ \ \ \ +54\la^4\vp^4 G^6+81\la^4\vp^4 G^6\right)
\eea
where $G \equiv1/(1+3\la\vp^2)$, and where we have defined 
$\Sigma^{(i)}\equiv \sum_{0\leq j\leq i} \Ga^{(j)}$.
The results for $\Ga_{\rm exact}$ and $\Sigma^{(i)}$ for three exemplary 
values of $\la$ and $\vp$ are given in the following table:
\beq
\begin{array}{|c|r@{.}l|r@{.}l|r@{.}l|r@{.}l|}
\hline
\la&1&0&1&0&100&0&200&0\\ \hline
\vp&1&0&4&0&1&0&1&0\\
\hline
\Ga_{\rm exact}&1&4532&73&9458145683&28&353282939&53&6991599696\\
\hline
S_{\rm cl}&0&75&72&0&25&5&50&5\\
\hline
\Sigma^{(1)}&1&4431&73&9459101483&28&353555132&53&6992974659\\
\hline
\Sigma^{(2)}&1&4431&73&9458145253&28&353282864&53&6915996015\\
\hline
\Sigma^{(3)}&1&4512&73&9458145667&28&353282912&53&6991599663\\
\hline
\end{array}\nn
\eeq
Evidently the approximation converges rapidly even for large values of $\la\,$!


\end{document}